\documentclass[twocolumn,showpacs,preprintnumbers,amsmath,amssymb,prl]{revtex4-1}
\usepackage{mathrsfs}
\usepackage{graphicx}
\usepackage{dcolumn}
\usepackage{bm}
\usepackage{amsmath}
\usepackage{amsfonts}
\usepackage{color}

\begin{document}

\title{Electrically Tunable Magnetism in Magnetic Topological Insulators}
\author{Jing Wang}
\author{Biao Lian}
\author{Shou-Cheng Zhang}
\affiliation{Department of Physics, McCullough Building, Stanford University, Stanford, California 94305-4045, USA}

\begin{abstract}
The external controllability of the magnetic properties in topological insulators would be important both for fundamental and practical interests. Here we predict the electric-field control of ferromagnetism in a thin film of insulating magnetic topological insulators. The decrease of band inversion by the application of electric fields results in a reduction of magnetic susceptibility, and hence in the modification of magnetism. Remarkably, the electric field could even induce the magnetic quantum phase transition from ferromagnetism to paramagnetism. We further propose a transistor device in which the dissipationless charge transport of chiral edge states is controlled by an electric field. In particular, the field-controlled ferromagnetism in magnetic topological insulator can be used for voltage based writing of magnetic random access memories in magnetic tunnel junctions. The simultaneous electrical control of magnetic order and chiral edge transport in such devices may lead to electronic and spintronic applications for topological insulators.
\end{abstract}

\date{\today}


\maketitle

\paragraph{Introduction}

The recent discovery of the quantum anomalous Hall (QAH) effect has attracted
intense interest in condensed matter physics~\cite{qi2011,hasan2010,qi2008,liu2008,li2010,yu2010,ruegg2011,wang2013a,wang2013b,onoda2003,chang2013b,kou2014,checkelsky2014,garrity2014,chang2015}.
The QAH insulator is a new state of quantum matter which has a
topologically nontrivial electronic structure characterized
by a bulk energy gap but gapless chiral edge states, leading
to the quantized Hall effect without an external magnetic
field~\cite{haldane1988}. The edge channels of the QAH insulator conduct without
dissipation, and thus has promising potential in the applications of low-power-consumption electronic devices.
The QAH effect has been observed
in thin films of Cr-doped~\cite{chang2013b,kou2014,checkelsky2014} and V-doped~\cite{chang2015} (Bi,Sb)$_2$Te$_3$ magnetic topological
insulators (MTIs), where robust bulk ferromagnetic (FM) ordering is
spontaneously developed in this system even in the insulating regime.

The ability to external control the magnetic properties of TIs~\cite{checkelsky2012,kou2013,zhang2014} could be important both for fundamental and technological interest, particularly in view of recent developments in magnetoelectrics and spintronics~\cite{zutic2004,fert2008}. In dilute FM semiconductors, the FM is mediated by itinerant charge carriers~\cite{jungwirth2006} and the magnetic ordering can be tuned by controlling the carrier density through an electric field~\cite{dietl2014}. But the electrical manipulation of magnetism in \emph{insulating} MTIs has proved elusive. Recently, the magnetic ordering in TIs is shown to be related to the band topology~\cite{zhang2013}, where the inverted band structure contributes a sizable Van Vleck magnetic susceptibility~\cite{yu2010}. Here we propose the electric-field control of FM in an insulating MTI thin film. The band inversion is weakened by applying an electric field, leading to a reduction of the magnetic susceptibility, which is directly related to the magnetism. Remarkably, the electric field could even induce a quantum phase transition (QPT) from the FM phase to the paramagnetic (PM) phases. The thin film MTI with strong FM exhibits the QAH effect with chiral edge states. Based on this property, we further propose a transistor device in which the dissipationless charge transport of topological edge states is controlled by an electric field.

\paragraph{Model}

We begin with introducing the topological properties of a generic two-dimensional (2D) MTI thin film.
For concreteness, we study the magnetically doped (Bi,Sb)$_2$Te$_3$ family materials. The low energy physics of the system is described by the Dirac-type surface states (SSs) only~\cite{yu2010,zhang2011,wang2014a}. The 2D effective Hamiltonian is
\begin{eqnarray}\label{model0}
\mathcal{H}_0(\vec{k})
   &=&\epsilon(\vec{k})+v_F k_y\sigma_1\otimes\tau_3-v_F k_x\sigma_2\otimes\tau_3
   \nonumber\\
   &&+m(\vec{k})1\otimes\tau_1+V1\otimes\tau_3,
\end{eqnarray}
with the basis of $|u\uparrow\rangle$, $|u\downarrow\rangle$, $|l\uparrow\rangle$ and $|l\downarrow\rangle$, where $u$, $l$ denote the upper and lower SSs and $\uparrow$, $\downarrow$ represent spin up and down states, respectively. For simplicity, we ignore the particle-hole
asymmetry term $\epsilon(\vec{k})$. $\vec{k}=(k_x,k_y)$. $\sigma_i$ and $\tau_i$ ($i=1,2,3$) are Pauli matrices acting on spin and layer, respectively. $v_F$ is the Fermi velocity. $V$ denotes the structure inversion asymmetry (SIA) between the two surfaces, which may come from band-bending induced by the substrate~\cite{zhang2010}, and can be tuned by applying an electric field along $z$ direction. $m(\vec{k})=m_0+m_1(k_x^2+k_y^2)$, describes the tunneling effect between the upper and lower SSs. For thick films $m_0=0$, the two SSs form gapless Dirac cones. In this case, it is suggested that these gapless Dirac SSs would mediate exchange coupling between magnetic moments through Ruderman-Kittel-Kasuya-Yoshida (RKKY) interaction~\cite{jungwirth2006,dietl2014}, leading to FM~\cite{liu2009}. For thin films $m_0\neq0$, and the coupling between these two SSs induces a hybridization gap. When the Fermi level is in the hybridized gap, the itinerant carrier mediated RKKY interaction is absent, and the Van Vleck mechanism instead takes over.

\begin{figure}[t]
\begin{center}
\includegraphics[width=2.6in]{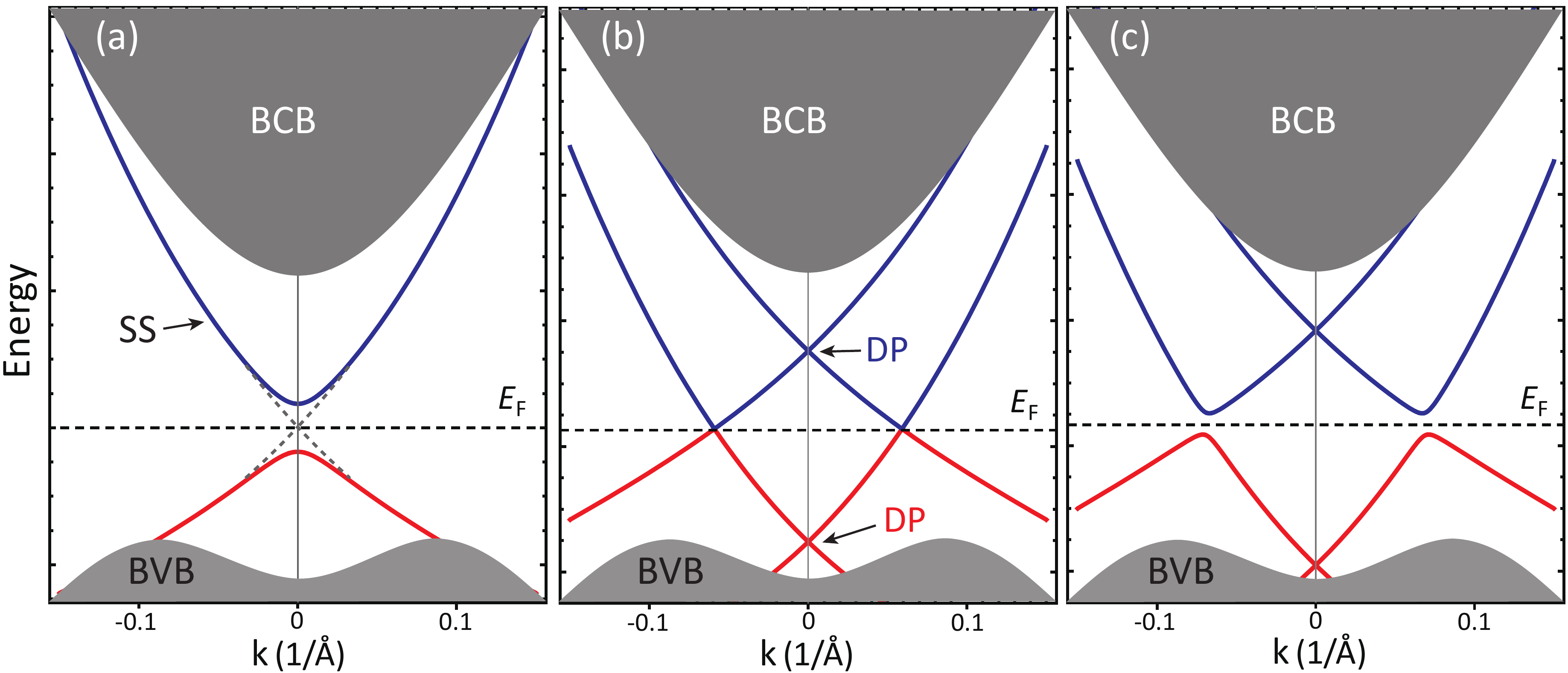}
\end{center}
\caption{(color online). Evolution of surface band structure upon increasing $V$ in TI thin films. The dashed line indicates the Fermi level $E_F$. BCB/BVB, bulk conduction/valence band; DP, Dirac point. (a) $V=0$, upper and lower SSs are degenerate with a hybridization gap opened at Dirac point. (b) $V=V^c_T$, upper and lower SSs move up and down in energy, respectively, and touch at a finite momentum $k_c=V^c_T/v_F$. (c) $V>V^c_T$, a band gap is reopened at $k_c$.}
\label{fig1}
\end{figure}

We consider in this Letter the $m_0\neq0$ case, where the band topology significantly affects the magnetism of the system~\cite{zhang2013} through the Van Vleck mechanism. In terms of the new basis $|+\uparrow\rangle$, $|-\downarrow\rangle$, $|+\downarrow\rangle$, $|-\uparrow\rangle$ with $|\pm\uparrow\rangle=(|u\uparrow\rangle\pm|l\uparrow\rangle)/\sqrt{2}$
and $|\pm\downarrow\rangle=(|u\downarrow\rangle\pm|l\downarrow\rangle)/\sqrt{2}$, the effective model becomes
\begin{eqnarray}\label{model1}
\widetilde{\mathcal{H}}_0(k_x,k_y) &=&
\begin{pmatrix}
\widetilde{\mathcal{H}}_+(k) & V\sigma_1\\
V\sigma_1 & \widetilde{\mathcal{H}}_-(k)
\end{pmatrix}.
\end{eqnarray}
Here, $\widetilde{\mathcal{H}}_{\pm}(k)=v_F\left(k_y\sigma_1\mp k_x\sigma_2\right)+m(k)\sigma_3$. If $V=0$, this model is similar to the Bernevig-Hughes-Zhang model for HgTe quantum wells~\cite{bernevig2006c}. When $m_0m_1>0$, the system is a normal insulator (NI) with $Z_2$ index $\nu=0$. When $m_0m_1<0$, the system is a quantum spin Hall (QSH) insulator with $\nu=1$. If $V\neq0$, such SIA term may induce the topological QPT from QSH to NI~\cite{shan2010}. The phase boundary is determined by the bulk gap closing. The energy spectrum is $E_0(k)=\pm\sqrt{(v_Fk\mp V)^2+m^2(k)}$, and the critical point is determined by $m(k)=0$ and $v_Fk=V$ which leads to critical $V^c_T=v_F\sqrt{-m_0/m_1}$. Here $V^c_T$ is the critical value of $V$ for the topological QPT from QSH to NI. The evolution of the band structure upon increasing the SIA $V$ is shown in Fig.~\ref{fig1}. For $V<V^c_T$, the system is adiabatically connected to the QSH state with a full gap. For $V>V^c_T$, the system is a NI.

\begin{figure}[b]
\begin{center}
\includegraphics[width=2.7in]{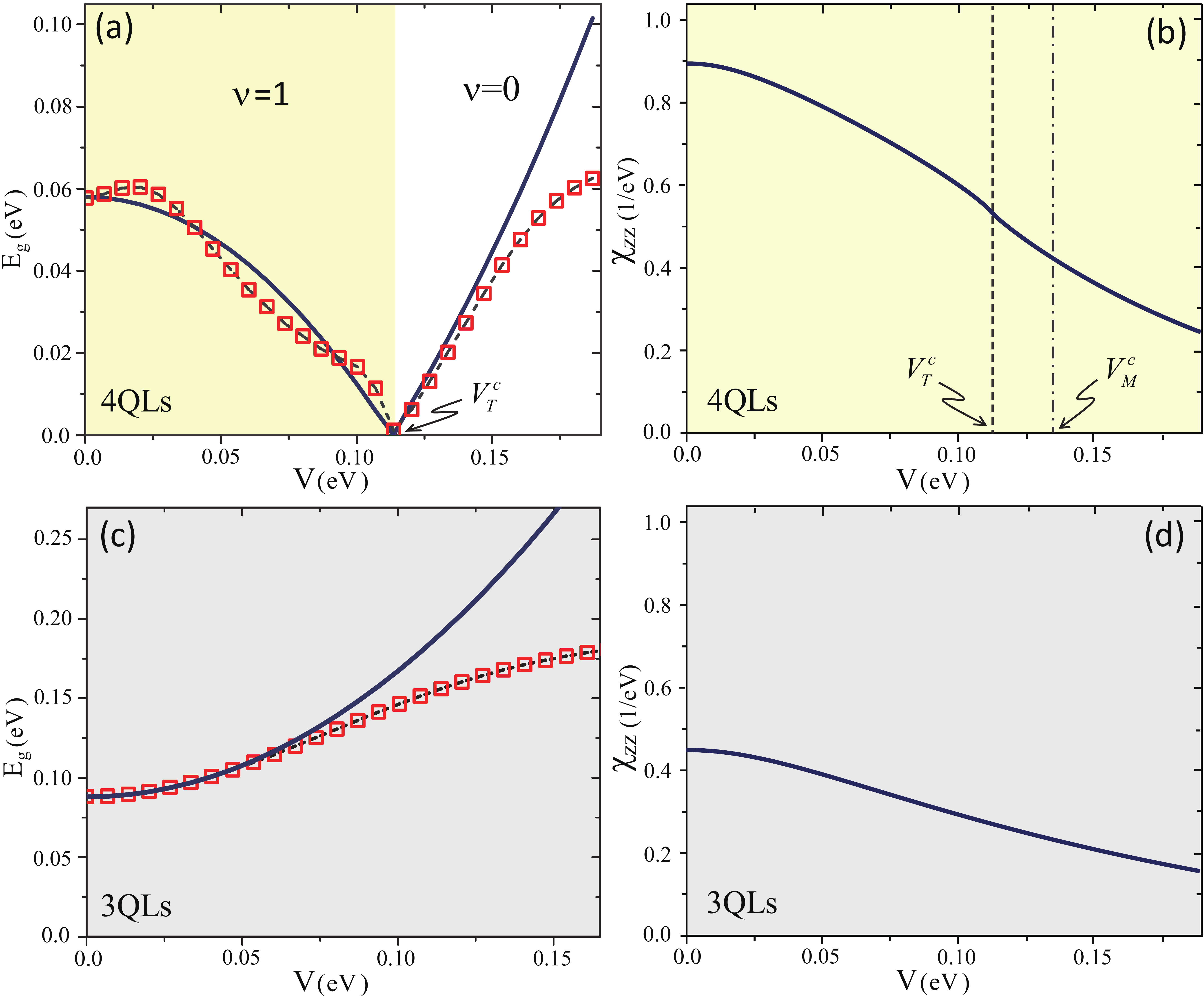}
\end{center}
\caption{(color online). Band structure and magnetic properties. (a), (c) Band gap of 4~QLs and 3~QLs
TIs, respectively, as a function of SIA $V$, calculated
analytically from Eq.~(\ref{model0}) (solid line) or numerically (symbols).
The parameters are taken from Ref.~\cite{zhang2009} for
(Bi$_{0.1}$Sb$_{0.9}$)$_{2}$Te$_{3}$. In (a), when $0\leq V<V^c_T$,
the band structure is inverted, resulting in a QSH with $Z_2$
index $\nu=1$; $V>V^c_T$, the band structure becomes normal. (b), (d) The calculated
spin susceptibility vs $V$ for 4~QLs and 3~QLs TIs, respectively. $V^c_M\approx0.13$~eV~\cite{supplementary}.}
\label{fig2}
\end{figure}

The above discussion based on the effective model gives us a clear physical picture of the
topological QPT driven by SIA. To confirm the validity of the picture
and estimate the magnitude of $V$, we calculate the band structure of thin film (Bi$_{0.1}$Sb$_{0.9}$)$_2$Te$_3$ in an external electric field along $z$ direction. We consider the three-dimensional (3D) bulk Hamiltonian $\mathcal{H}_{\mathrm{3D}}$~\cite{zhang2009}
in a thin film configuration with thickness $d$, where the effect of an external electric field is modeled by adding
$V_E(z)=\mathcal{E}z/d$. The confinement in the $z$ direction quantizes the momentum on
this axis and leads to 2D subbands. We solve the eigen equation $[\mathcal{H}_{\mathrm{3D}}+V_E(z)]\psi_{n\vec{k}}(z)=E_{n\vec{k}}\psi_{n\vec{k}}(z)$
with open boundary condition $\psi_{n\vec{k}}(0)=\psi_{n\vec{k}}(d)=0$, where $n$ is the subband index.
By projecting the bulk model onto the lowest four subbands,
the parameters in Eq.~(\ref{model0}) for different quintuple layers (QLs) can be obtained, as shown in Table~\ref{table1}.
Each QL is about 1~nm thick. The calculated band gap for different QLs are shown in Fig.~\ref{fig2}.
Good agreement between the numerical calculation and analytic model is found, which confirms the validity of the effective model in
Eq.~(\ref{model0}). The discrepancy between them at large $V$ suggests low energy physics is no longer
dominated by Dirac-type SSs only when $V$ is comparable to the subbands splitting. The energy of bulk subbands shifts little at moderate $V$. The system oscillates between QSH insulator and NI as
a function of layer thickness~\cite{liu2010a,supplementary}. The enhanced $V$ with increasing film thickness shows the efficient
tunability of band structure by an electric field.

\begin{table}[t]
\caption{The parameters of the 2D effective Hamiltonian in Eq.~(\ref{model0}) for 3 and 4 QLs (Bi$_{0.1}$Sb$_{0.9}$)$_2$Te$_3$ thin films.}
\begin{center}\label{table1}
\renewcommand{\arraystretch}{1.2}
\begin{tabular*}{3.3in}
{@{\extracolsep{\fill}}ccccc}
\hline
\hline
& $v_F$ (eV \AA) & $m_0$ (eV) & $m_1$ (eV \AA$^2$) & $V$ (eV)
\\
\hline
4QLs & $2.36$ & $-0.029$ & $12.9$ & $0.212\mathcal{E}$
\\
3QLs & $3.07$ & $+0.044$ & $37.3$ & $0.134\mathcal{E}$
\\
\hline
\hline
\end{tabular*}
\end{center}
\end{table}

\paragraph{Susceptibility $\&$ FM}

The magnetic properties of MTIs is determined by the effective interaction between
localized magnetic impurity spins $-\mathcal{J}_{\mathrm{eff}}\vec{S}_I(\vec{r}_i)\cdot\vec{S}_I(\vec{r}_j)$, where $\vec{S}_I(\vec{r}_i)$ denotes the magnetic impurity spin at the position $\vec{r}_i$. There are two mechanisms contributed to $\mathcal{J}_{\mathrm{eff}}$.
The first mechanism corresponds to a virtual process where both electrons (holes) on two impurities hop into the itinerant bands, have their spins correlate with each other and then hop back. The strength of such interaction is
given by $-\mathcal{J}_{F}=-J^2\chi_e$. Here $J$ is the exchange coupling parameter between the magnetic moments and itinerant
electron spin described by $\mathcal{H}_{\mathrm{ex}}=J\sum_{\vec{r}_i}\vec{S}_I(\vec{r}_i)\cdot\vec{s}$, $\vec{s}=\vec{\sigma}/2$
is the spin operator, and $\chi_e$ is the spin susceptibility of electrons. For temperatures much lower than the band gap, the Van Vleck type spin susceptibility for a band insulator is
\begin{equation}
\chi_e^{zz} = \sum\limits_{\vec{k},m,n}\mathrm{Tr}\left(s_zP_{m\vec{k}}s_zP_{n\vec{k}}\right)\frac{f(E_{n\vec{k}})-f(E_{m\vec{k}})}{E_{m\vec{k}}-E_{n\vec{k}}},
\end{equation}
where $f$ is the Fermi-Dirac distribution function,
$E_{n\vec{k}}$ is the energy of $n$-th subband, and the projection operator $P_{n\vec{k}}=|\psi_{n\vec{k}}\rangle\langle\psi_{n\vec{k}}|$.
As shown in Fig.~\ref{fig2}, we calculated the $z$-direction spin susceptibility ($\chi_{e}^{zz}$) of both 4~QLs and 3~QLs TI films as a function of $V$,
where the chemical potential is always in the gap. For 4~QLs, $\chi^{zz}_e$ remains a large value in the inverted regime,
while is reduced sufficiently in the non-inverted regime by increasing $V$. For 3~QLs, $\chi^{zz}_e$ also decreases as $V$ increases, but much slower.
Therefore, the interaction $\mathcal{J}_{F}$ is electrically tunable. Here the susceptibility tensor is anisotropic with $\chi_e^{zz}/\chi_e^{xx}\approx1.7$, such anisotropy is also reduced as $V$ increases~\cite{supplementary}.
The second mechanism is superexchange,
where the electron (hole) on one impurity virtually hops to the others via the itinerant bands and then hops back in the same way.
The strength of superexchange interaction $\mathcal{J}_{A}\sim J^2/U$, where $U$ is the Hubbard interaction energy on impurity.
Such superexchange coupling is antiferromagnetic (AFM) with $\mathcal{J}_A>0$, which depends little on the itinerant band features, and
barely changes under an electric field.

Now the effective interaction $\mathcal{J}_{\mathrm{eff}}$ is given by
$\mathcal{J}_{\mathrm{eff}} = \mathcal{J}_F-\mathcal{J}_A$.
The spin susceptibility of the local moment then takes the Curie-Weiss form $\chi=C/(T-T_c)$. $C=\mu_J^2/3k_B$,
$\mu_J$ is the magnetic moment of a single magnetic impurity. $T_c$ is the Curie temperature of FM, which
reads $T_c=x\mathcal{J}_{\mathrm{eff}}/k_B$ according to the mean field theory. Here we have assumed the Ising like bahavior of magnetic impurities spin. $x$ is the concentration of magnetic impurities. Take $x=3\%$ for example, $T_c\approx15$~K for Cr-doped TI~\cite{chang2013b}, while $T_c\approx60$~K for V-doped TI~\cite{chang2015}.
The electrically tunable $\mathcal{J}_F$ and constant $\mathcal{J}_A$ lead to a remarkable consequence: the decrease of band inversion by the application of electric field results in a reduction of magnetic susceptibility and magnetic anisotropy, and hence in the modification of magnetism ($\mathcal{J}_F\rightarrow\mathcal{J}_{\mathrm{eff}}\rightarrow T_c$, $H_c$). The reduced $\mathcal{J}_{\mathrm{eff}}$ and magnetic anisotropy with increasing $V$ would result in the reduction of the coercive field $H_c$~\cite{supplementary}.
More interestingly, the electric field could even induce the magnetic QPT from FM-to-PM when $\chi^{zz}_e<1/U$ (where $T_c<0$ means the system is no longer FM). Here PM refers to phases with PM response~\cite{PM_note}, such phases may include PM, AFM and spin glass~\cite{AFM_note}. The exact phase depends in detail on the dopant distributions and distance dependence of magnetic interaction, which has not been settled down yet in experiments~\cite{bao2013,hesjedal2014,patankar2015}. For an estimation, taking $U=2.5$~eV for the $d$ orbitals of $3d$ transition metals such as Cr, the FM-to-PM transition happen when $\chi^{zz}_{e}\sim0.4/\mathrm{eV}$. When $V>V^c_M$, $\chi^{zz}_e<1/U$. $V^c_M$ is the critical value of $V$ for the magnetic QPT. With $0.2$~mm SrTiO$_3$ as the dielectric substrate and $40$~nm Al$_2$O$_3$ as the top dielectric as shown in Fig.~\ref{fig4}(a), $V_{\mathrm{bg}}=50$~V and $V_{\mathrm{tg}}=20$~V are needed to drive such FM-to-PM transition~\cite{supplementary}.
Moreover, the value of $\chi^{zz}_e$ in 3~QLs is smaller than that in 4~QLs, due to non-inverted mass and larger gap, which naturally explains weaker FM order in 3~QLs compared to that in 4~QLs~\cite{feng2015}. We emphasize that the topological and magnetic QPTs discussed above are different. However, the topological QPT and the associated band inversion will change of the magnetic susceptibility and magnetic anisotropy sufficiently, making the observation of magnetic QPT feasible.

\paragraph{QAH $\&$ phase diagram}

\begin{figure}[b]
\begin{center}
\includegraphics[width=2.2in]{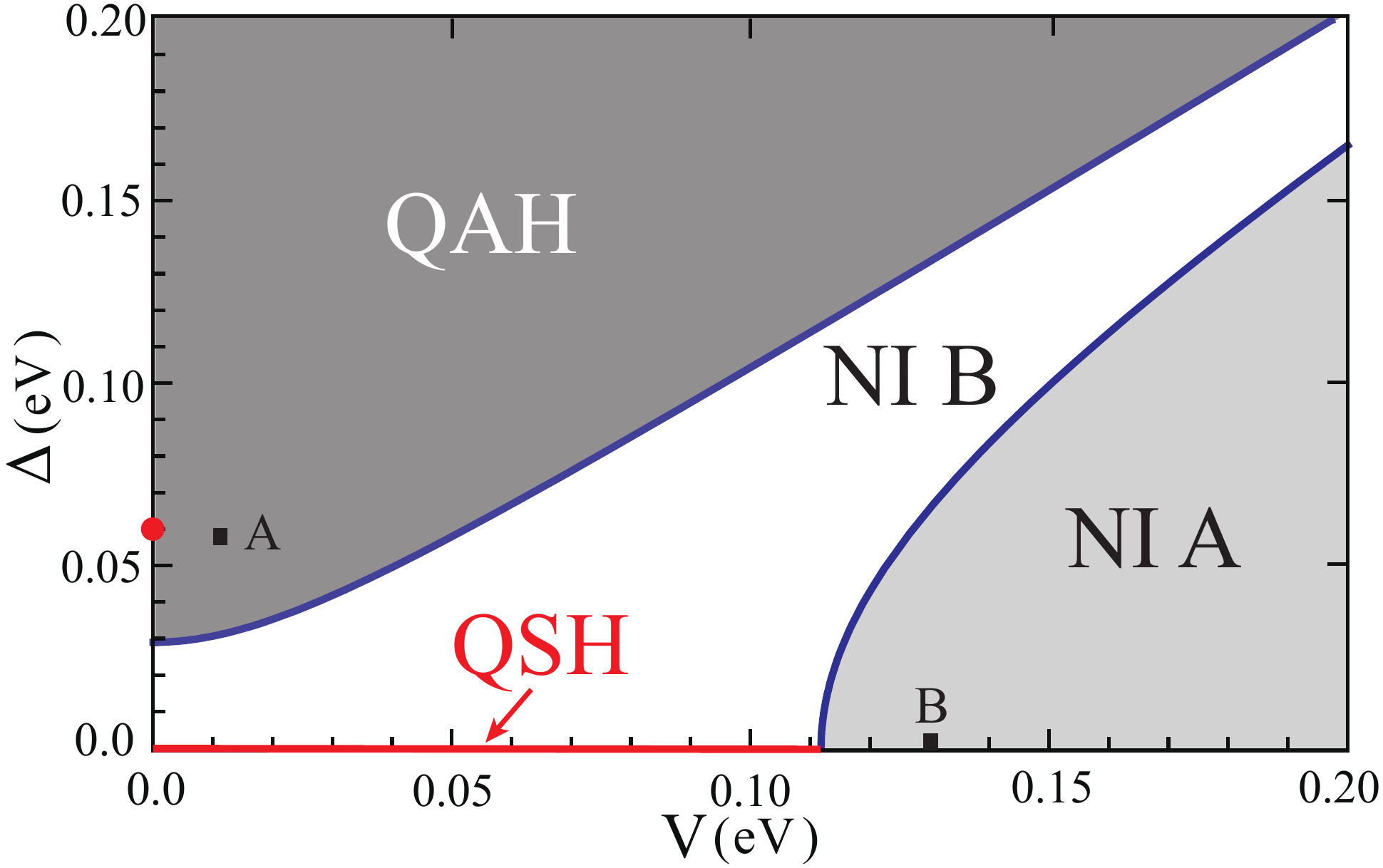}
\end{center}
\caption{(color online). The phase diagram of the thin films of MTIs for $m_0\neq0$ with two variables:
$\Delta$ and $V$. Only $\Delta\geq0$ and $V\geq0$ is shown.
The parameters are taken from Table~\ref{table1} for 4~QLs Cr$_{0.15}$(Bi$_{0.1}$Sb$_{0.9}$)$_{1.85}$Te$_{3}$.
Phase QSH is well defined only in the $\Delta=0$ line. The phase boundary (navy solid lines) is semi-metal phase.
The edge spectrum of points A and B are shown in Figs. 4(b) and 4(c), respectively.}
\label{fig3}
\end{figure}

Below $T_c$, the FM order is developed. The magnetic moments align and induce a
Zeeman type spin splitting $\Delta$ for the SSs due to exchange coupling along $z$ axis. The mean field effective Hamiltonian is,
$\mathcal{H}_{\mathrm{MF}}=\mathcal{H}_0+\Delta\sigma_3\otimes1$.
In the absence of SIA $V=0$, the system will be in the QAH phase as long as $|\Delta|>|m_0|$~\cite{wang2014a}. In realistic materials,
however, such condition is hard to achieve in 3~QLs due to weak FM
order, thus 3~QLs is a NI. With strong FM order, QAH state is realized in 4~QLs~\cite{feng2015}, where estimated $\Delta\approx0.06$~eV~\cite{supplementary} (showed as the red dot in Fig.~\ref{fig3}).
We further consider the $V\neq0$ case.
Similarly, we determine the phase boundaries by the gapless regions in the energy spectrum,
which leads to:
(i), $m_0^2+V^2 = \Delta^2$; or (ii), $m(k) = 0$ and $\Delta^2+v_F^2k^2=V^2$.
The entire phase diagram for 4~QLs in the $(\Delta,V)$ space is shown in Fig.~\ref{fig3}.
Except for the phase boundaries, there are four gapped phases. NI A is a band insulator with zero
charge Hall conductance, while NI B is a band insulator with both
spin and charge Hall conductance be nonzero and non-quantized.
As predicted, the SIA would induce the FM-to-PM transition, therefore $\Delta$ decreases as $V$ increases,
and without the FM order, the system would be a NI. Therefore, the electrical control of FM will drive
the QAH-to-NI phase transition, where the phase trajectory in real materials would follow a curve from QAH to NI A in Fig.~\ref{fig3}.
\begin{figure}[b]
\begin{center}
\includegraphics[width=2.2in]{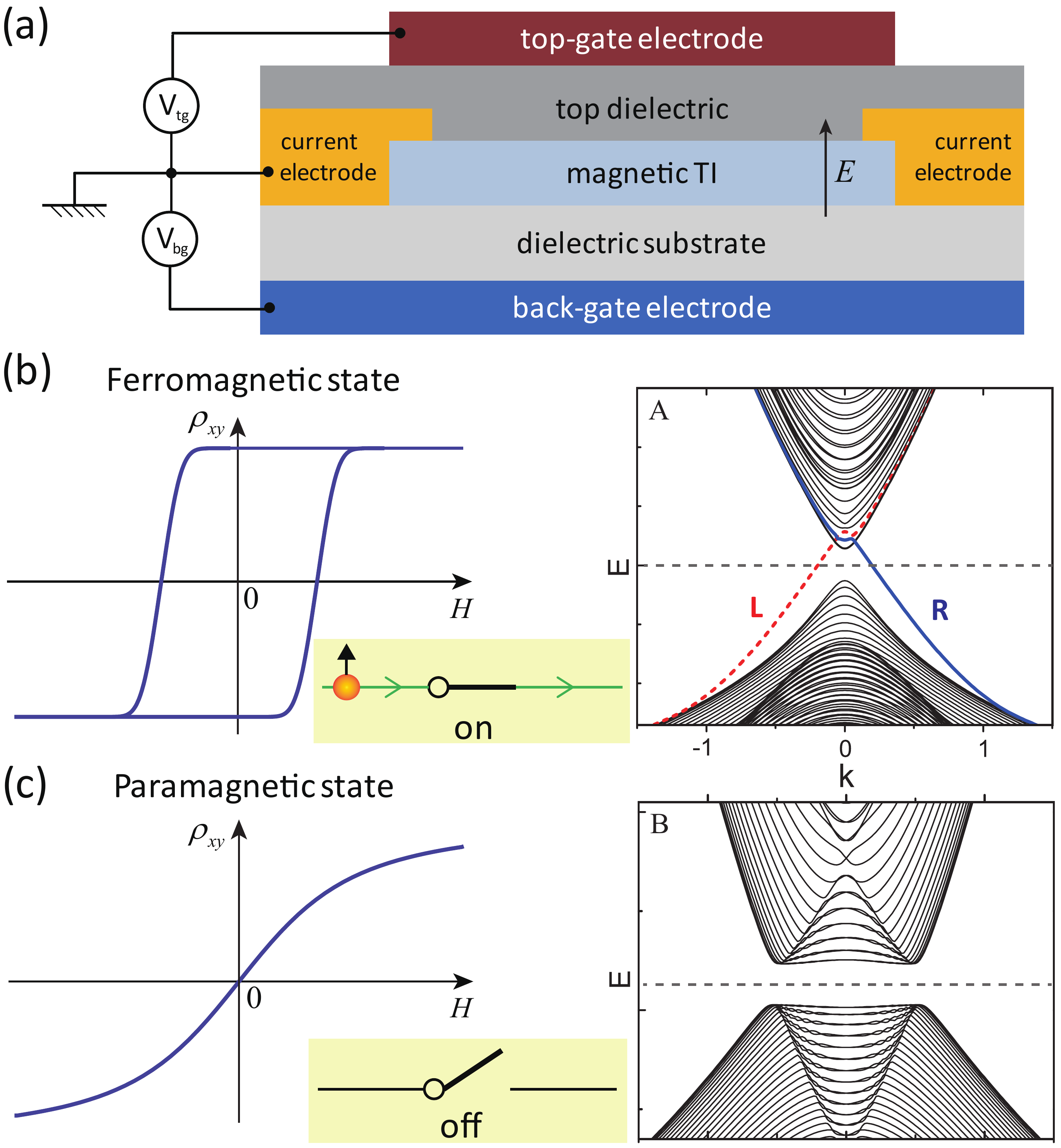}
\end{center}
\caption{(color online). (a) Schematic diagram of a proposed transistor device for using dual gates and an electric field to tune charge/spin transport. $V_{\text{tg}}$, top-gate voltage; $V_{\text{bg}}$, back-gate voltage.
(b) Without an electric field, the MTI film has FM order and is in a QAH phase, thus protected, chiral edge state (upper right). (c) Applying an electric field perpendicular to the film breaks the inversion symmetry, which drives the MTI to be PM and in a NI phase, thus no protected edge state (lower right).}
\label{fig4}
\end{figure}

\paragraph{Transistor device proposal}

The metallic chiral edge states conduct without dissipation in the QAH phase;
while these states disappear in NI. Thus, the conductance by the edge state transport in 2D MTI is electrically tunable instead of carrier depletion, which improves power efficiency and can work at high on/off ratio~\cite{onoff_note}. In the on state, the current is carried by the \emph{dissipationless} edge states. In the off state, only a local
electric field is needed, which minimizes power consumption. Based on this, we propose a transistor device made of dual-gated MTI thin films as shown in Fig.~\ref{fig4}(a). Using the two gates, one can control the electric field across the film and SIA, thus turn on and off the charge transport by purely electrical means.

\begin{figure}[t]
\begin{center}
\includegraphics[width=2.6in]{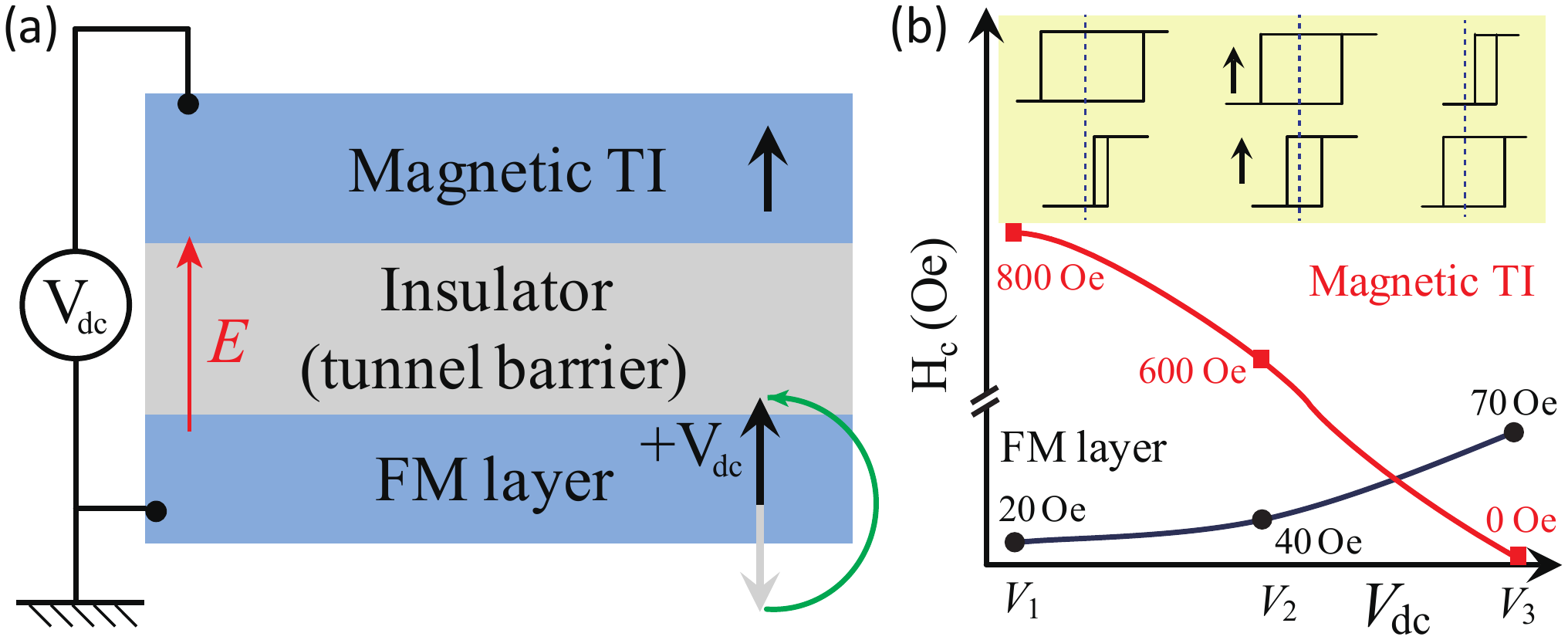}
\end{center}
\caption{(color online). Electric-field-assisted switching in a MTJ, which can be used as voltage based writing of MRAM. (a) Schematic drawing of a MTJ and the effect of electric field. (b) Dependence of $H_c$ for the top MTI and bottom FM layer on the bias voltage. The estimation of $H_c$ and working condition of the device are discussed in SI~\cite{supplementary}.}
\label{fig5}
\end{figure}

\paragraph{MRAM proposal}

The $H_c$ of MTIs decreases as $V$ increases, and remains FM as along as $V<V^c_M$~\cite{supplementary}. Such property of electrical manipulation of $H_c$ can be used for voltage based writing of non-volatile magnetic random access memories (MRAM) in the magnetic tunnel junctions (MTJs). As proposed in Fig.~\ref{fig5}, the MTI is the top FM layer, MgO is the insulator barrier layer, and CoFeB is the bottom FM layer with out-of-plane magnetic anisotropy. It has been shown that $H_c$ of bottom CoFeB layer with certain thickness increases with increasing $V$, which is due to enhanced perpendicular magnetic anisotropy with depleting electrons at CoFeB/MgO interface~\cite{wangwg2012}. With such device configuration,
one can realize the electric-field-assisted reversible switching of FM in the MTJs, here a small bias
magnetic field is needed in the setup. More details on the working condition for such device are presented in Supplemental Material~\cite{supplementary}. Compared with the spin transfer torque effect, the magnetic configuration and magnetic tunneling magnetoresistance in MTJs can be manipulated by voltage pulses with much smaller current densities.

\paragraph{Conclusion}

We predict the field-controlled FM in MTIs, which is expected to have a great impact for electronic and spintronic applications of TIs. Such prediction is generic for TI materials close to the topological QPT.
We emphasize that the modulation of $T_c$ in ferromagnet (In,Mn)As by an electric field is due the electric control of carrier concentration~\cite{ohno2000}, while the modulation of $T_c$ in MTIs predicted here by an electric field is due to the electric control of band inversion and spin texture of band structures.

\begin{acknowledgments}
This work is supported by the US Department of Energy, Office of Basic Energy Sciences, Division of Materials Sciences and Engineering, under Contract No.~DE-AC02-76SF00515 and the Defense Advanced Research Projects Agency Microsystems Technology Office, MesoDynamic Architecture Program (MESO) through the Contract No.~N66001-11-1-4105, and in part by FAME, one of six centers of STARnet, a Semiconductor Research Corporation program sponsored by MARCO and DARPA.
\end{acknowledgments}


\begin{thebibliography}{45}
\bibitem [{AFM()}]{AFM_note}%
  \BibitemOpen
  \href@noop {} {}\bibinfo {note} {When $\chi^{zz}_e<1/U$, the AFM
  superexchange interaction prevails. However, in dilute magnetically doped
  (Bi,Sb)$_2$Te$_3$ TIs, Bi/Sb forms triangle lattice and the magnetic dopants
  randomly replace Bi/Sb in the system. Such random distribution of magnetic
  dopants and triangle lattice act as frustration. Therefore, the system may be
  an AFM state, spin glass or PM.}\BibitemShut {Stop}%
\bibitem [{ono()}]{onoff_note}%
  \BibitemOpen
  \href@noop {} {}\bibinfo {note} {The resistance of the off state depends on
  the amount of impurity states inside the energy gap, and thus the on/off
  ratio can be improved by controlling the film quality. On the other hand,
  even in the diffusive transport regime, the on/off ratio of the transistor
  can be further enhanced with a quantized on state conductance of $e^2/h$ per
  edge and a negligible off state conduction.}\BibitemShut {Stop}%
\end{thebibliography}
\end{document}